**@AGU PUBLICATIONS**

# Journal of Geophysical Research: Space Physics

**RESEARCH ARTICLE**
10.1002/2014JA020882

# Distribution of energetic oxygen and hydrogen in the near-Earth plasma sheet

E. A. Kronberg[1], E. E. Grigorenko[2], S. E. Haaland[1,3], P. W. Daly[1], D. C. Delcourt[4], H. Luo[5], L. M. Kistler[6], and I. Dandouras[7,8]

[1]Max Planck Institute for Solar System Research, Göttingen, Germany, [2]Space Research Institute, Russian Academy of Sciences, Moscow, Russia, [3]Birkeland Centre for Space Science, Department of Physics, University of Bergen, Bergen, Norway, [4]Laboratoire de Physique des Plasmas, Palaiseau Cedex, France, [5]Key Laboratory of Ionospheric Environment, Institute of Geology and Geophysics, Chinese Academy of Sciences, Beijing, China, [6]Department of Physics Space Science Center, University of New Hampshire, Durham, New Hampshire, USA, [7]University of Toulouse, UPS-OMP, UMR 5277, Institut de Recherche en Astrophysique et Planátologie, Toulouse, France, [8]CNRS, IRAP, Toulouse, France

**Abstract** The spatial distributions of different ion species are useful indicators for plasma sheet dynamics. In this statistical study based on 7 years of Cluster observations, we establish the spatial distributions of oxygen ions and protons at energies from 274 to 955 keV, depending on geomagnetic and solar wind (SW) conditions. Compared with protons, the distribution of energetic oxygen has stronger dawn-dusk asymmetry in response to changes in the geomagnetic activity. When the interplanetary magnetic field (IMF) is directed southward, the oxygen ions show significant acceleration in the tail plasma sheet. Changes in the SW dynamic pressure ($P_{dyn}$) affect the oxygen and proton intensities in the same way. The energetic protons show significant intensity increases at the near-Earth duskside during disturbed geomagnetic conditions, enhanced SW $P_{dyn}$, and southward IMF, implying there location of effective inductive acceleration mechanisms and a strong duskward drift due to the increase of the magnetic field gradient in the near-Earth tail. Higher losses of energetic ions are observed in the dayside plasma sheet under disturbed geomagnetic conditions and enhanced SW $P_{dyn}$. These observations are in agreement with theoretical models.

## 1. Introduction

The distribution of charged particles in the near-Earth plasma sheet has been discussed in various aspects in previous studies. *Meng et al.* [1981], using energetic proton (50–500 keV) observations from 5 years (each) of IMP 7 and 8 data, at ~$30R_E < R <~ 40R_E$, reported dawn-dusk asymmetry in the plasma sheet with intensities of protons higher at the duskside. *Sarafopoulos et al.* [2001] studied the asymmetry of dawn-dusk plasma sheet energetic particles at <25 to 850 keV in the ~15–28 $R_E$ downtail plasma sheet from the Interball tail probe. They reported duskward asymmetry for ions as well. The asymmetry was found for the average oxygen energy by *Ohtani et al.* [2011] at distances < $15R_E$, for oxygen densities at the dayside by *Bouhram et al.* [2005], in MHD simulations by *Winglee and Harnett* [2011] for relative oxygen energy density and in simulations by *Fok et al.* [2006] based on the Lyon-Fedder-Mobarry MHD model [e.g., *Fedder et al.*, 1995] for the plasma pressure. However, no or very small asymmetries in the densities of protons and oxygen were observed by *Mouikis et al.* [2010] (0–40 keV/e), *Ohtani et al.* [2011] (9–210 keV/e) (also for the average proton energy), and *Maggiolo and Kistler* [2014] (0–40 keV/e). Therefore, there still is no clear, consistent picture of the spatial distribution of ions and their dependence on magnetospheric disturbances in the magnetosphere [*Kronberg et al.*, 2014]. The underlying physics that creates these distributions is also poorly understood.

An acceleration by quasi-stationary dawn-dusk electric fields alone (like Speiser acceleration [*Speiser*, 1965]) cannot accelerate ions to energies higher than the typical tail potential drop (not more than 100 keV). Also, pure betatron acceleration (as reported, e.g., by *Sarafopoulos et al.* [2001]) cannot lead to these energies. Therefore, the ions at energies >100 keV must be accelerated by a different mechanism. Induced electric fields can accelerate ions to energies well exceeding the typical value of the potential drop across the tail (higher than 100 keV). Induction electric fields [see, e.g., *Delcourt*, 2002] can result from the fast magnetic X line formation [*Zelenyi et al.*, 1990], electromagnetic turbulence [*Grigorenko et al.*, 2011], or current disruption processes [*Lui*, 1996; *Lutsenko et al.*, 2008]. *Nosé et al.* [2000] using 3 years of Geotail/EPIC measurements of energetic (60 keV to 3.6 MeV) ions reported that, in substorms, the energetic particle flux of $O^+$ ions is more enhanced than that of H$^+$ ions in the near-Earth tail ($X \sim -16$ to $-6R_E$). They stated that the strong increase of energetic oxygen ions was due to the local magnetic field reconfiguration (dipolarization) and not due to magnetotail reconnection. Effective acceleration of ions up to energies higher than $\sim$140 keV, in the vicinity of the near-Earth X line, was statistically demonstrated by *Luo et al.* [2014].

Accelerated plasma sheet particles drift earthward and populate the ring current which influences the formation of the radiation belts. Energetic ions at energies from 274 to 955 keV may contribute to the dynamics of the ring current which is responsible for the disturbances of the terrestrial magnetic field, especially during magnetic storms [*Kozyra and Liemohn*, 2003; *Ganushkina et al.*, 2005]. The rapid increase of oxygen pressure in the nightside ring current at substorm expansion was revealed, e.g., in case study by *Mitchell et al.* [2005], simulations by *Fok et al.* [2006], and reviewed by *Keika et al.* [2013]. Various simulations have shown that the inclusion of the energetic oxygen ions is crucial for reproducing the ring current and for the magnetospheric dynamics, in general (as at the same energy with protons it has 4 times higher energy density and therefore pressure) [*Glocer et al.*, 2009; *Fok et al.*, 2011; *Winglee and Harnett*, 2011]. Ions at energies higher than thermal also serve as a seed population for formation of the radiation belts.

Variations of the plasma pressure define not only the growth but also decay of the ring current. This can lead to the dawn-dusk asymmetry at the dayside and can be, e.g., due to the losses of ions at the magnetopause. The open drift paths lead to escape at the dayside magnetopause [e.g., *Paschmann*, 1997; *Keika et al.*, 2005; *Wang et al.*, 2013]. It is interesting to check the presence of asymmetry in the spatial distributions of energetic ions of different masses under various geomagnetic and interplanetary conditions.

In this study we investigate how the spatial distribution of the energetic oxygen and protons from 274 to 955 keV in the plasma sheet depends on the SW dynamics and geomagnetic activity. Ion distribution patterns can give us a hint on the acceleration, transport, and losses in the plasma sheet and to assess simulation studies. Studying the ions at energies >274 keV, we can find out under which SW and geomagnetic conditions the inductive acceleration mechanisms are the most effective and whether it is different for oxygen and protons. We present a comprehensive study of the distribution of energetic oxygen and hydrogen ion abundances in the near-Earth plasma sheet (from $-20\ R_E < X_{GSM} < 10\ R_E$). The measurements are obtained from 7 years of particle measurements from the Cluster satellites. The novelty of this study is a combination of (a) the extensive region coverage of the near-Earth magnetosphere dayside and mainly toward the flanks at the nightside, as previous studies either focused on the magnetotail region or were case studies in the near-Earth region; (b) energy range, as no extensive statistical study has been done so far for energies up to 955 keV; (c) for the first time we thoroughly look at how the spatial distribution of these populations depends on the geomagnetic activity and SW dynamics from the tail to the near-Earth plasma sheet at the dayside; and (d) we compare our results with recent numerical models.

The paper is organized as follows. In section 2, we give a brief overview of the data set and describe how the data maps were constructed. Section 3 shows how the spatial distribution of the oxygen and hydrogen ions in the near-Earth magnetosphere varies for different levels of geomagnetic and SW disturbance. Section 4 discusses the results and related physical processes. Section 5 summarizes the results.

## 2. Instrumentation, Data, and Methodology

The results presented in this study are primarily based on in situ measurements from the Cluster spacecraft for years 2001–2007. More information about the Cluster mission and instrumentation is given in *Escoubet et al.* [1997]. In this paper we used 1 min averaged omnidirectional energetic ion intensities from spacecraft 4, since this gives the best data return for our purpose.

We utilized data from the "Research with Adaptive Particle Imaging Detector" (RAPID) [*Wilken et al.*, 2001] taking the combined energy channels from 274 to $\sim$955 keV. This energy range is chosen as this is the lowest range for the oxygen ion measurements by RAPID instrument [*Daly and Kronberg*, 2010]. The method on how those channels were created and further details on related data processing can be found in *Kronberg et al.* [2012].

### 2.1. Construction of Maps

In our study the plasma sheet region is defined by plasma beta values in the range 0.2–10 [*Baumjohann et al.*, 1989; *Grigorenko et al.*, 2012]. The plasma beta is calculated using Cluster Ion Spectrometry (CIS)/COmposition and DIstribution Function (CODIF) plasma pressure observations [*Rème et al.*, 2001] and the magnetic field observations by fluxgate magnetometer [*Balogh et al.*, 2001]. Median values of the plasma beta (fluctuating around 1) for different geomagnetic and SW conditions are shown in Table 1.

**Table 1.** Median Values of the *Dst* Index, the *AE* Index, SW $P_{dyn}$, $IMF_{B_z}$, the Plasma Beta, the Number of Records, and the Median Number of Records Per Bin for Oxygen in Figures 2–4

| Type of Map | *Dst* (nT) | *AE* (nT) | SW $P_{dyn}$ (nPa) | $IMF_{B_z}$ (nT) | Beta | Number of Records | Number Per Bin |
|---|---|---|---|---|---|---|---|
| $AE \leq 150$ nT | −13.5 | 64 | 0.97 | 1.4 | 1.04 | 17,315 | 77 |
| $AE \geq 250$ nT | −30.5 | 505 | 2.04 | −2.2 | 0.86 | 26,560 | 113 |
| SW $P_{dyn} \leq 1.5$ nPa | −21 | 177 | 0.52 | −0.72 | 0.90 | 24,803 | 85 |
| SW $P_{dyn} \geq 2$ nPa | −23 | 391 | 3.21 | −0.67 | 1.00 | 20,325 | 112 |
| $IMF_{B_z} \geq 2$ nT | −17 | 133 | 1.83 | 4.19 | 0.97 | 14,025 | 70 |
| $IMF_{B_z} \leq -2$ nT | −29 | 452 | 1.54 | −5.16 | 0.86 | 18,977 | 76 |

Using this beta range, we mostly include the plasma sheet ions [*Grigorenko et al.*, 2012]. The amount of ions from the plasma sheet boundary layers (PSBL) is at most 30%. In order to totally exclude PSBL ions one should use the beta ≥5. At those beta values it is impossible to get meaningful statistics for the maps. We have investigated how the maps will change if we use beta ≥1 (in this case less than 17% ions can be from PSBL [*Grigorenko et al.*, 2012]). We found no substantial difference in ion distributions. Additionally we controlled distributions using another criteria for the plasma sheet $B_x/B < 0.8$. These distributions also do not show substantial differences in this case. Therefore, for the better coverage we choose to show maps without two latter restrictions.

We project the ion intensities into the $XY_{GSM}$ plane and do not map them along field lines, as we are interested in the content of the ions (proxy of energy density, which is an approximation of plasma pressure) in the plasma sheet as it is. In addition the considered ions have large gyroradius (e.g., the gyroradius of oxygen ion, taking 511 keV as the geometric mean of the considered energy channel, is about $2R_E$ considering the averaged magnetic field over this data set $\simeq 25$ nT) and mapping along magnetic field lines would be rather uncertain.

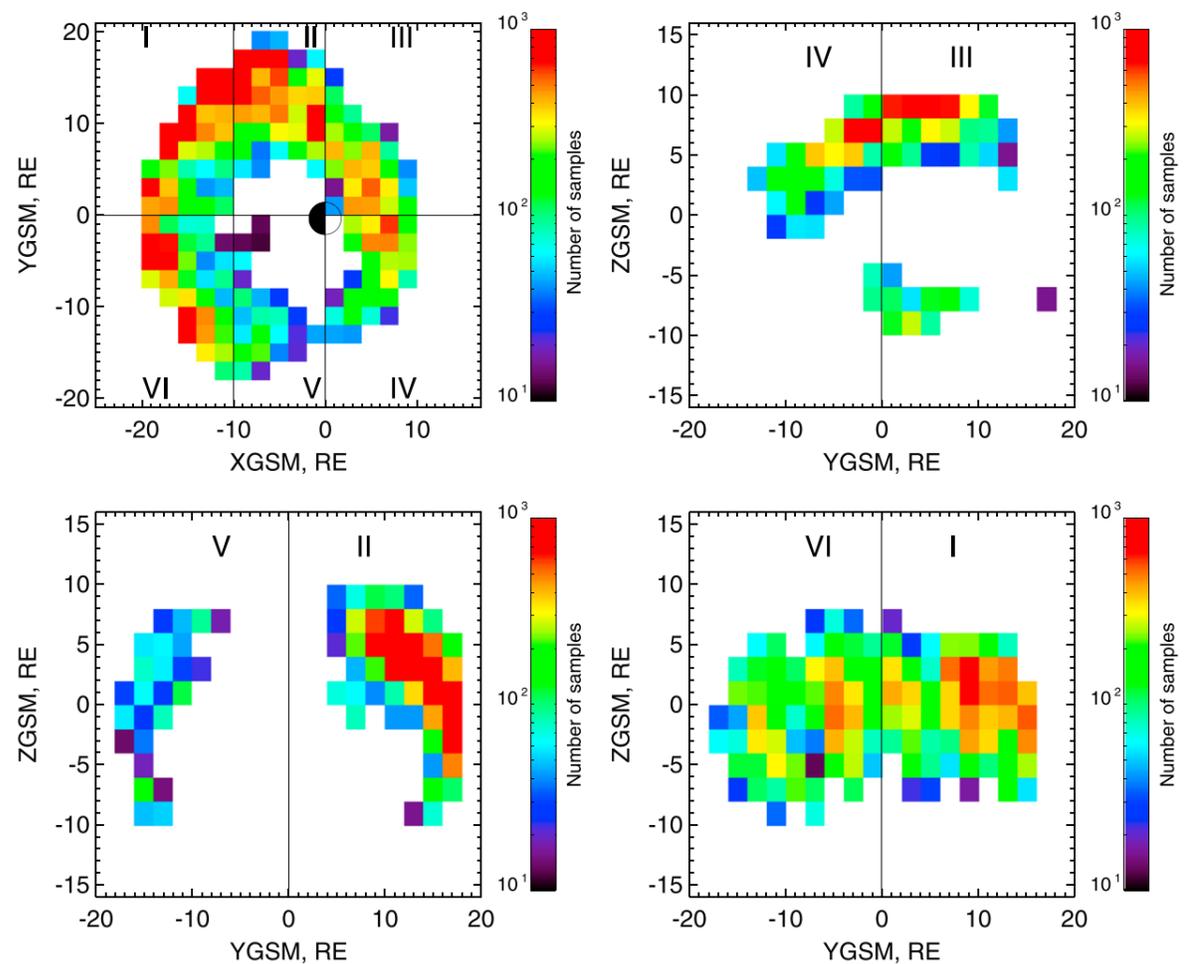

**Figure 1.** Maps of the sample number of the 1 min averaged energetic oxygen observations, (top left) in $XY_{GSM}$ (dayside and nightside) and (top right) in $YZ_{GSM}$ dayside (all data for $0 < X_{GSM}$); (bottom left) in $YZ_{GSM}$ nightside, near-Earth (all data for $-10 < X_{GSM} \leq 0\ R_E$) and (bottom right) $YZ_{GSM}$ nightside, tail (all data for $-10 < X_{GSM} \leq 0\ R_E$) for plasma beta between 0.2 and 10 and $-9 \geq Z_{GSM} \leq 9\ R_E$.

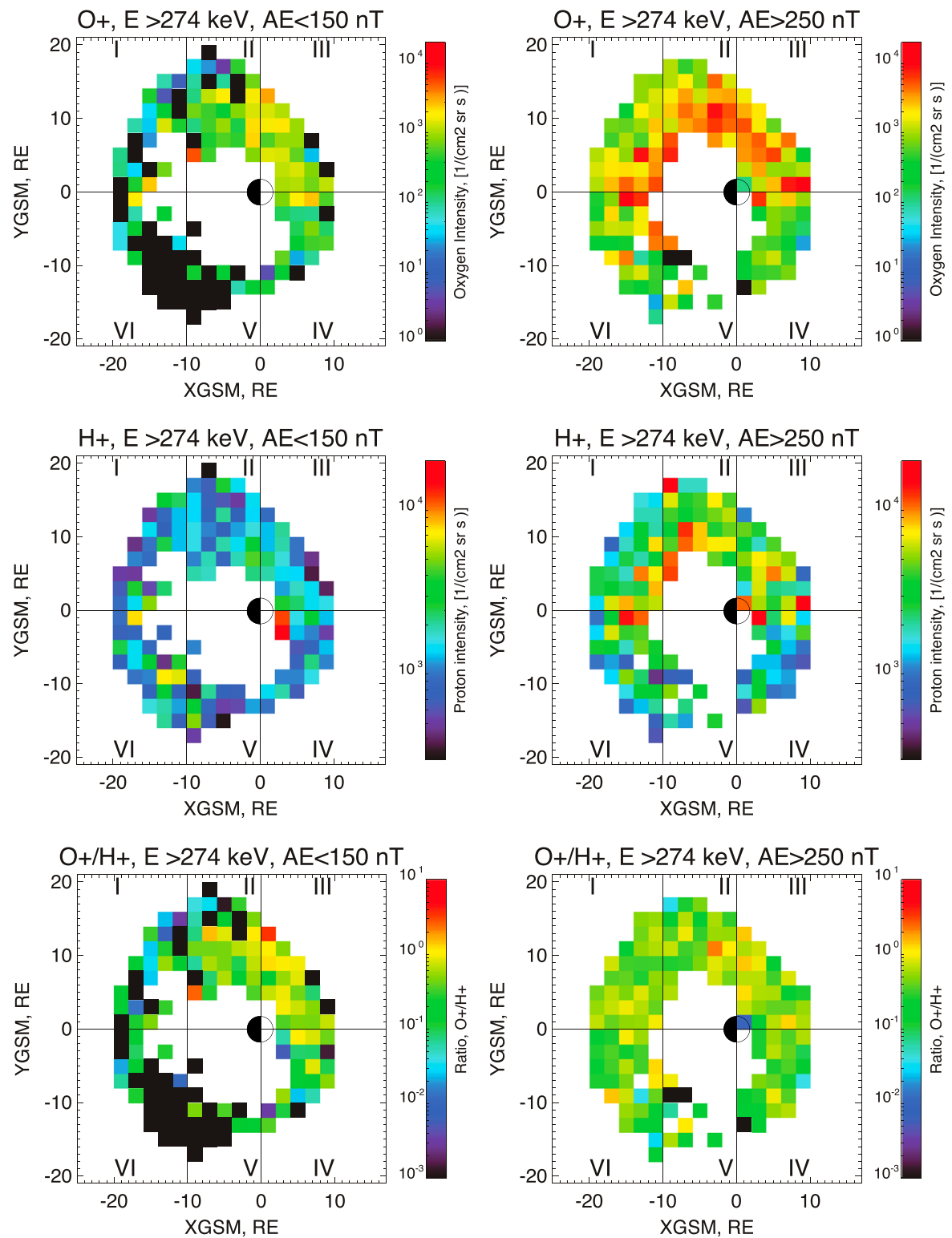

**Figure 2.** Maps of ion intensities at $E > 274$ keV versus $AE$ index, with (left) $AE \leq 150$ nT, (right) $AE \geq 300$ nT. (top) $O^+$, (middle) $H^+$, and (bottom) $O^+/H^+$. Lower/higher than in color bar values of intensities are presented by the minimum/maximum value in the color bar. Please note that the parts I and VI are more filled with data during disturbed times. This is not because there are more ions but that there were fewer measurements here during quiet times. The size of the Earth is not to scale.

The spatial coverage of >274 keV oxygen ion measurements in the geocentric solar magnetospheric (GSM) coordinate system used in this study is shown in Figure 1. As a result of spacecraft trajectories in the near-Earth region our data mainly cover the dawn and dusk flanks in the tail and the dayside plasma sheet, primarily in the north. This coverage allows us to study the transport of ions from the tail region around the Earth. We also restricted the vertical extend of observations to $|Z_{GSM}| \leq 9$ in order to keep the coverage more symmetric. The asymmetry of the spatial coverage which is still present at the dayside along $Z_{GSM}$ axis (see Figure 1, top right) does not affect our conclusions. For example, the dawn-dusk asymmetry at the dayside for energetic ions is still present if we take more narrow $Z_{GSM}$ ranges (to keep approximately the same latitude, not shown here). Radial distances are chosen to be greater than $R > 6\,R_E$ to avoid possible contaminations in radiation belts.

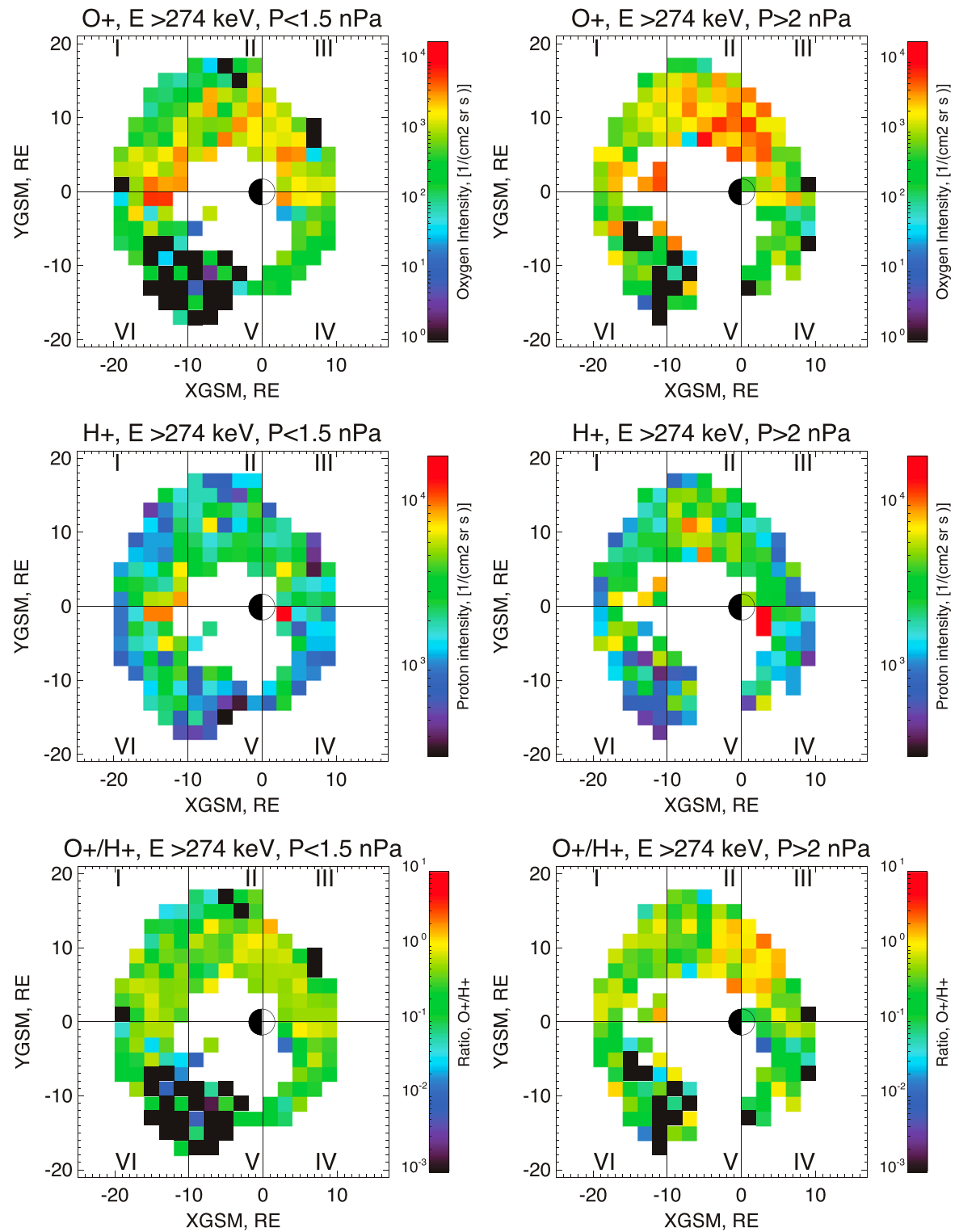

**Figure 3.** Maps of ion intensities at $E > 274$ keV versus the SW $P_{dyn}$, with (left) $P_{dyn} \leq 1.5$ nT, (right) $P_{dyn} \geq 2$ nT. In the same format as Figure 2.

Ion intensity maps for energies >274 keV for two geomagnetic activity levels, quiet ($AE \leq 150$ nT) and disturbed ($AE \geq 250$ nT) are shown in Figure 2, for two SW dynamic pressure ($P_{dyn}$) levels, low ($P \leq 1.5$ nPa) and high ($P \geq 2$ nPa) in Figure 3, and for two IMF directions, northward ($IMF_{B_z} \geq 2$) and southward ($IMF_{B_z} \leq -2$) in Figure 4. As a compromise between resolution and statistics, we use $2R_E \times 2R_E$ bins, and only plot values where each bin has more than 10 data records from three independent Cluster orbits. White color shows where there is a lack of data, and black color shows where the measurements yield zero counts. If a data point has a value equal to zero or less than the minimum color bar value, it is painted in black color. We have divided the maps into six regions (I–VI) for better comparison. The total number of 1 min averaged records and the median number of samples per $2R_E \times 2R_E$ bin for each type of map are shown in Table 1. We show the numbers only for oxygen as the difference from those for protons is less than 0.1%. Additionally we show the median values of the $Dst$ index, the $AE$ index, SW $P_{dyn}$, and $IMF_{B_z}$ for each type of map in Table 1. To minimize the effect of skewed distribution, we use median rather than mean values of the fluxes for construction of the maps, see details in *Kronberg et al.* [2012].

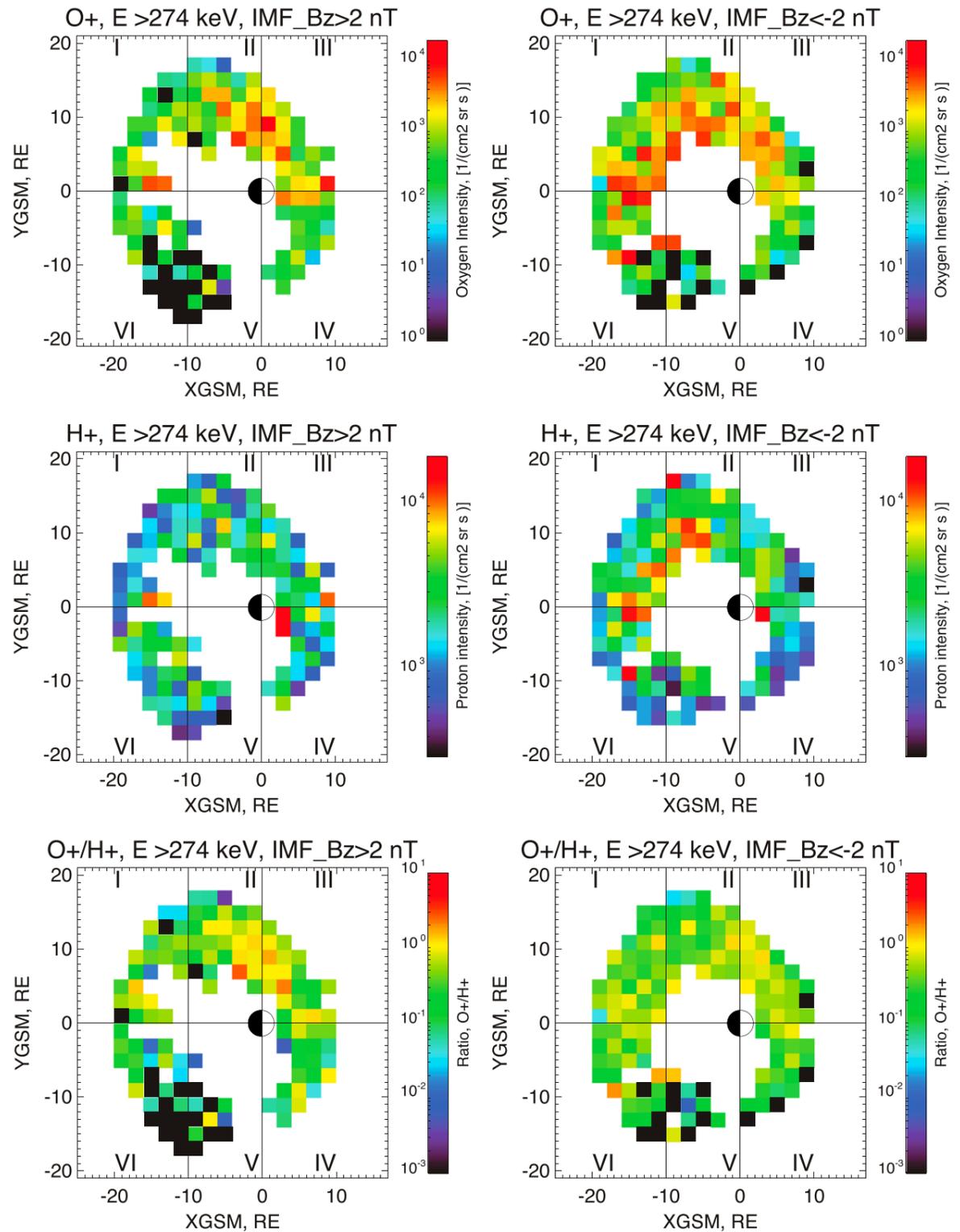

**Figure 4.** Maps of ion intensities at $E > 274$ keV versus IMF $B_z$ direction, with (left) IMF $B_z \geq 2$ nT, (right) IMF $B_z \leq -2$ nT. In the same format as Figure 2.

In Table 1, one can see that the low (high) SW $P_{dyn}$ and the northward (southward) IMF$_{B_z}$ are correlated with lower (higher) *AE* index. We checked the linear Pearson correlation between these parameters in our database: *AE* versus SW $P_{dyn}$ is −0.008 and *AE* versus IMF$_{B_z}$ is −0.11 (for the test, SW pressure versus SW density is 0.99). This means that *AE* index is quite independent from the SW $P_{dyn}$ and the IMF$_{B_z}$. We have also tried to plot the ion distributions for the SW $P_{dyn}$ and the IMF$_{B_z}$ at specific *AE* ranges in order to eliminate the apparent dependence on *AE*. However, in such cases we do not have enough points of measurements to do meaningful statistics.

We checked if seasonal effects related to Cluster orbit affect our conclusions. Namely, the regions I and II are always traversed by Cluster during approximately September–November, when the IMF-magnetosphere coupling is most efficient. We compared the median *AE* index for dusk and dawn observations at the nightside during both quiet and disturbed times. The numbers are the following: 75 nT and 56 nT for quiet times, 528 nT and 522 nT for disturbed times, for dawn and dusk, respectively. This difference should not significantly affect our conclusions.

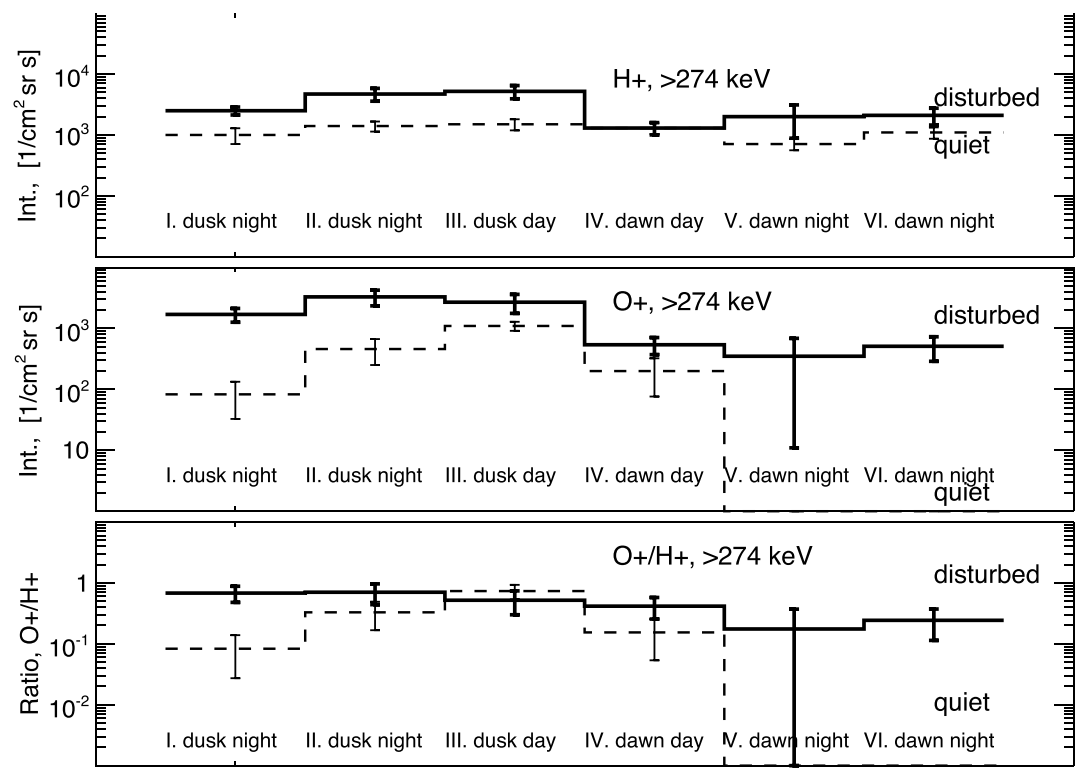

**Figure 5.** Dependencies on local time of (first panel) proton intensities at >274 keV, (second panel) oxygen intensities at >274 keV, and (third panel) ratio of $O^+/H^+$ at >274 keV, derived from Figure 2. The thick solid lines show the intensities for the disturbed times ($AE \geq 250$ nT) and thin dashed ones for the quiet times ($AE \leq 150$ nT). Error bars indicate 95% confidence intervals.

According to the results of earlier statistical studies there is a 1–2 h delay between the southward turn of the IMF and the generation of the accelerated ion beams in the current sheet sources [*Borovsky et al.*, 1998; *Grigorenko et al.*, 2005]. The SW $P_{dyn}$ may affect the distribution in the plasma sheet with a time delay. We have plotted the ion distributions with two different time delays (0.5 h and 1.5 h), finding no significant difference in comparison with the distributions without delay. This means that the delays vary quite a lot depending on the particular acceleration source which accelerates the ions to high energies (e.g., magnetic reconnection, dipolarizations, and so on). Several acceleration sources may simultaneously operate in the magnetotail. As a result, what we see in the plasma sheet is an "average" distribution of energetic ions reflecting the global dependence on the IMF and SW conditions.

In order to better visualize and quantify the dawn-dusk asymmetries Figures 5–7 show the distributions of ion intensities and their ratios in the dawn and dusk regions, on the dayside and nightside, also dividing nightside into two regions for two geomagnetic activities, SW $P_{dyn}$ and IMF $B_z$ levels, respectively. The numbers and error bars used in these figures are listed in Table 2. The intensity values are derived by taking the median value of the intensity of all $2R_E \times 2R_E$ bins in the corresponding region. By this, we lose information on magnetic local time and radial variations. In order to calculate error bars in Figures 5–7, we first calculated the median absolute deviation which we then converted to standard deviation using a factor of 1.4826 [*Huber*, 1981] (see more details on how the statistical analysis was done in *Kronberg et al.* [2012]). In order to simplify comparison of the values, the confidence interval (CI) error bars (CI = $t_{n-1} \cdot \sigma/\sqrt{n}$, where $t_{n-1}$ is the Student's $t$ distribution with $n$ degrees of freedom and $\sigma$ the standard deviation) are calculated in Figures 5–7. The Student's coefficient $t_{n-1}$ is used at the 5% level of significance. Two values are considered to be significantly different if the confidence interval error bars do not overlap each other. Two values do not show significant difference if the standard error bars (SE = $\sigma/\sqrt{n}$, not shown in figure but basically half of the confidence interval) overlap each other.

### 3. Results

We check here for prominent spatial asymmetries in the ion distributions and significant differences between different geomagnetic and SW conditions.

For the different levels of geomagnetic activity the main findings are the following:

**Table 2.** Median Values of the Ion Intensities (cm$^{-2}$sr$^{-1}$s$^{-1}$) in the Different Regions in Figures 2–4

| Species | Region I | Region II | Region III | Region IV | Region V | Region VI |
|---|---|---|---|---|---|---|
| | | | $AE \leq 150nT$ | | | |
| O$^+$ | 83 ± 50 | 460 ± 209 | 1100 ± 186 | 200 ± 124 | 0 ± 0 | 0 ± 0 |
| H$^+$ | 1000 ± 291 | 1400 ± 263 | 1500 ± 310 | 1300 ± 266 | 710 ± 150 | 1100 ± 233 |
| O$^+$/H$^+$ | 0.08 ± 0.06 | 0.33 ± 0.16 | 0.73 ± 0.20 | 0.15 ± 0.10 | 0 ± 0 | 0 ± 0 |
| | | | $AE \geq 250nT$ | | | |
| O$^+$ | 1700 ± 435 | 3300 ± 951 | 2700 ± 930 | 540 ± 169 | 350 ± 339 | 510 ± 221 |
| H$^+$ | 2500 ± 371 | 4700 ± 1119 | 5200 ± 1279 | 1300 ± 296 | 2000 ± 1115 | 2100 ± 648 |
| O$^+$/H$^+$ | 0.68 ± 0.20 | 0.70 ± 0.26 | 0.52 ± 0.22 | 0.42 ± 0.16 | 0.18 ± 0.20 | 0.24 ± 0.13 |
| | | | $SWP_{dyn}, P \leq 1.5nPa$ | | | |
| O$^+$ | 900 ± 344 | 860 ± 344 | 1500 ± 297 | 400 ± 118 | 0 ± 0 | 140 ± 73 |
| H$^+$ | 2000 ± 470 | 2100 ± 354 | 2300 ± 297 | 1400 ± 316 | 1100 ± 367 | 1900 ± 475 |
| O$^+$/H$^+$ | 0.45 ± 0.20 | 0.41 ± 0.18 | 0.65 ± 0.15 | 0.29 ± 0.11 | 0 ± 0 | 0.07 ± 0.04 |
| | | | $SWP_{dyn}, P \geq 2nPa$ | | | |
| O$^+$ | 1300 ± 365 | 2400 ± 540 | 1700 ± 774 | 400 ± 170 | 390 ± 378 | 190 ± 106 |
| H$^+$ | 2700 ± 513 | 4900 ± 755 | 2800 ± 824 | 1600 ± 303 | 1200 ± 872 | 1300 ± 425 |
| O$^+$/H$^+$ | 0.48 ± 0.16 | 0.49 ± 0.13 | 0.61 ± 0.33 | 0.25 ± 0.12 | 0.33 ± 0.39 | 0.15 ± 0.09 |
| | | | $IMF_{B_z} \geq 2nT$ | | | |
| O$^+$ | 540 ± 255 | 880 ± 432 | 1700 ± 533 | 440 ± 171 | 0 ± 0 | 74 ± 42 |
| H$^+$ | 1600 ± 297 | 2300 ± 432 | 2400 ± 571 | 1800 ± 634 | 1400 ± 696 | 2100 ± 570 |
| O$^+$/H$^+$ | 0.34 ± 0.17 | 0.38 ± 0.20 | 0.71 ± 0.28 | 0.24 ± 0.13 | 0 ± 0 | 0.04 ± 0.02 |
| | | | $IMF_{B_z} \leq -2nT$ | | | |
| O$^+$ | 1500 ± 584 | 1700 ± 637 | 1400 ± 667 | 490 ± 247 | 95 ± 80 | 480 ± 224 |
| H$^+$ | 2700 ± 531 | 4300 ± 934 | 2400 ± 545 | 1300 ± 390 | 1800 ± 1124 | 1700 ± 503 |
| O$^+$/H$^+$ | 0.56 ± 0.24 | 0.40 ± 0.17 | 0.58 ± 0.31 | 0.38 ± 0.22 | 0.05 ± 0.06 | 0.28 ± 0.16 |

1. The oxygen intensities are significantly higher at dusk than at dawn (about ∼8 times for quiet periods and ∼5.5 times during disturbed periods). There is a significant drop of the ion intensity between the postnoon and prenoon regions (ion intensity is ∼5 times lower during both quiet and disturbed times, compared regions III and IV), see Figures 2 and 5. The average intensity of >274 keV oxygen ions (for the current data coverage) shows an increase by a factor almost 5 during disturbed times. The most dramatic change in intensities between quiet and disturbed times, more than an order of magnitude is at the tail plasma sheet side (regions I and VI).

2. In contrast to >274 keV oxygen ions, the protons do not show significant asymmetry between dawn and dusk during quiet times (between regions II and VI, III and IV), as shown in Figures 2 and 5. For disturbed periods, the proton intensity is significantly higher on the duskside (dusk ion intensity is approximately 3 times of dawn ion intensity, compared regions II–V) but in the tail regions (I and VI) they do not show significant difference. At the dayside the intensity of protons drops by ∼4 times during disturbed time. The averaged intensity of >274 keV protons (for the current data coverage) during disturbed conditions is approximately 2.5 higher than during quiet periods.

3. The O$^+$/H$^+$ ratio of >274 keV ions is on average 0.21 during quiet times and 0.45 during disturbed times. This is significantly higher than the density ratios derived by *Maggiolo and Kistler* [2014]. This also means that if we would compare the energy densities which are proxies of the particle pressure, oxygen dominates the pressure at these energies in most of the regions covered here during disturbed times. The O$^+$/H$^+$ ratio and the oxygen intensity show almost identical distribution patterns, namely dawn-dusk asymmetry, during quiet time. There is a significant dawn-dusk asymmetry of the O$^+$/H$^+$ ratio in the tail region during disturbed times. This implies different acceleration mechanisms and/or type of particle sources for protons and oxygen ions. At the dayside the loss mechanism is similar for both oxygen ions and protons. However, during quiet time the relative loss of oxygen is more significant than of protons.

For the different levels of the SW $P_{dyn}$ the main findings are the following:

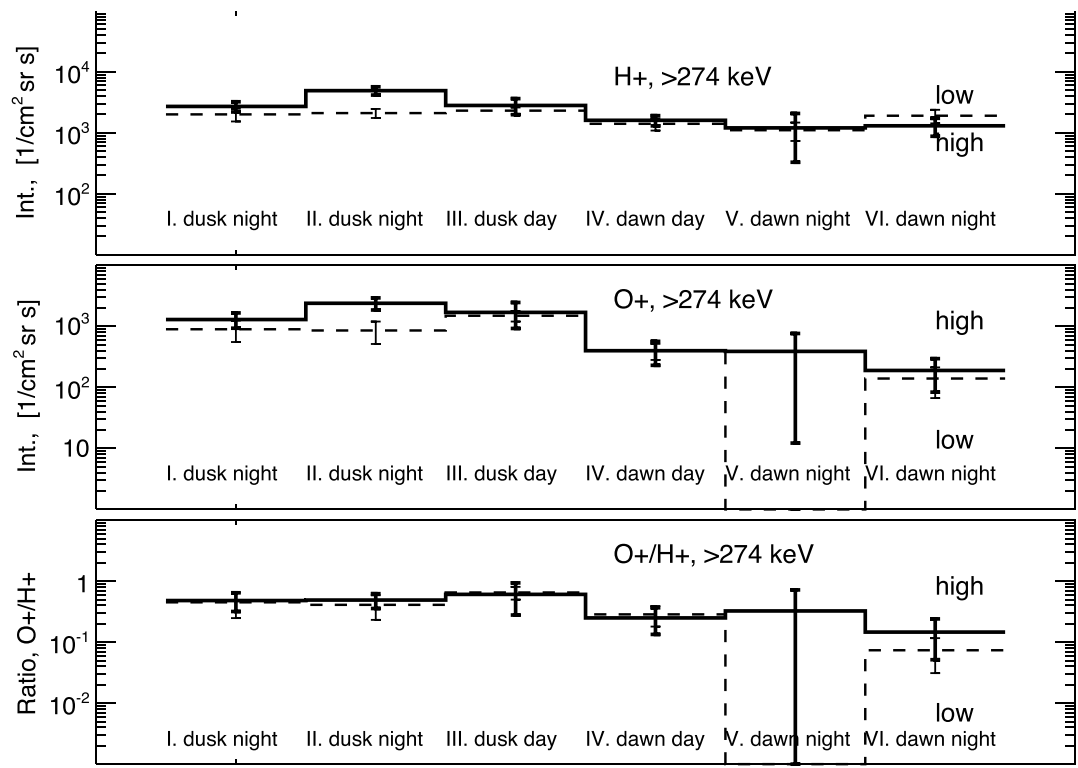

**Figure 6.** Dependencies on local time of (first panel) proton intensities at >274 keV, (second panel) oxygen intensities at >274 keV, and (third panel) ratio of $O^+/H^+$ at >274 keV. The thick solid lines show the intensities for the high SW $P_{dyn}$ (≥2 nPa) and thin dashed ones for the low SW $P_{dyn}$ (≤1.5 nPa), derived from Figure 3. Error bars indicate 95% confidence intervals.

4. The oxygen intensities are significantly higher at dusk than at dawn (about ~6 times at times of low SW $P_{dyn}$ and ~5.5 times at times of high SW $P_{dyn}$), see Figures 3 and 6. There is a significant drop of the ion intensity between the postnoon and prenoon regions during both low and high SW $P_{dyn}$ (ion intensity is ~4 times lower, compared regions III and IV). We do not observe an expected significant overall intensity increase from low to high SW $P_{dyn}$ (as flux tubes are compressed) but rather local increase, about 3 times, at the dusk near-Earth side (region II).

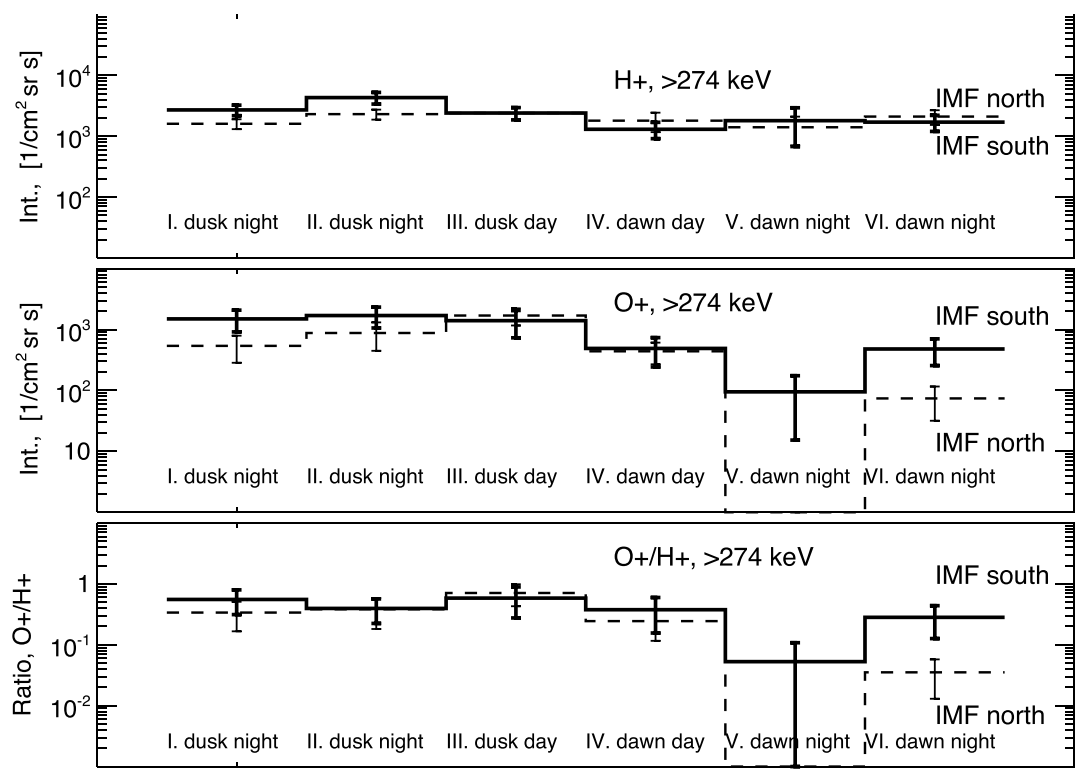

**Figure 7.** Dependencies on local time of (first panel) proton intensities at >274 keV, (second panel) oxygen intensities at >274 keV, and (third panel) ratio of $O^+/H^+$ at >274 keV. The thick solid lines show the intensities for the southward IMF (≤ −2 nT) and thin dashed ones for the northward IMF (≥2 nT), derived from Figure 4. Error bars indicate 95% confidence intervals.

Table 3. Presence of the Significant Dawn-Dusk Asymmetry Between Different Regions (Tail—regions I and VI, near-Earth nightside—regions II and V, dayside—regions III and IV) in Figures 2–4[a]

| Region | Oxygen | | Protons | | $O^+/H^+$ | |
|---|---|---|---|---|---|---|
| | Q | D | Q | D | Q | D |
| *AE Index* | | | | | | |
| Tail | yes | yes | no | no | yes | yes |
| Near-Earth nightside | yes | yes | yes | yes | yes | yes |
| Dayside | yes | yes | no | yes | yes | no |
| SD between "Q" and "D" | all regions excluding IV | | regions I–III | | regions I, IV, and VI | |
| *SW $P_{dyn}$* | | | | | | |
| Tail | yes | yes | no | yes | yes | yes |
| Near-Earth nightside | yes | yes | yes | yes | yes | no |
| Dayside | yes | yes | yes | yes | yes | no |
| SD between Q and D | region II | | region II | | no | |
| *$IMF_{B_z}$ Orientation* | | | | | | |
| Tail | yes | yes | no | no | yes | no |
| Near-Earth nightside | yes | yes | no | yes | yes | yes |
| Dayside | yes | yes | no | yes | yes | no |
| SD between Q and D | regions I, V, and VI | | regions I and II | | region VI | |
| *Kp Index for Ion Densities From Maggiolo and Kistler [2014]* | | | | | | |
| Tail | no | no | no | no | no | no |
| Near-Earth nightside | no | no | no | no | no | no |
| Dayside | no | no | no | no | no | no |
| SD between Q and D | all regions excluding IV | | region V | | regions I and III–VI | |

[a] Q and D mean "Quiet" and "Disturbed" conditions, respectively. For SW $P_{dyn}$ the higher pressure is Disturbed. For IMF orientation the southward direction is considered to be Disturbed. Additionally we show the regions where the significant difference (SD) between Quiet and Disturbed conditions is observed.

5. The dawn-dusk asymmetry of protons is less prominent than for oxygen ions during low SW $P_{dyn}$. However, the asymmetry becomes more prominent at times when the SW $P_{dyn}$ is high (2.5 times higher at the duskside). The same as for oxygen a significant intensity increase, about 2 times, is observed only at the dusk near-Earth side (region II) between the two SW $P_{dyn}$ levels.

6. There is no significant difference in the $O^+/H^+$ ratio between two levels of the SW $P_{dyn}$. This means that the SW $P_{dyn}$ affects oxygen and proton ions in a similar way, probably controlling ionospheric ion outflow [*Cully et al.,* 2003a] or the volume of the magnetosphere. Also, this means that SW $P_{dyn}$ does not affect the relative efficiency of proton and oxygen acceleration.

For the different directions of the IMF $B_z$ the main findings are the following:

7. The oxygen intensities are significantly higher on the duskside than on the dawnside (about ∼6 times for northward IMF and ∼4 times during southward IMF). There is a significant drop of the ion intensity at between the postnoon and prenoon regions (ion intensity is ∼4 times lower during northward IMF and ∼3 times during southward IMF, compared regions III and IV), see Figures 4 and 7. A significant change in intensities between northward and southward IMF, a factor of 3, is at the tail plasma sheet side (regions I and VI).

8. In contrast to >274 keV oxygen ions, the protons do not show significant asymmetry between dawn and dusk during northward IMF, as shown in Figures 4 and 7. However, they show significant dawn-dusk asymmetry during southward IMF (factor of 2 higher at the duskside, regions II–V). The asymmetry is observed at the dayside the proton intensity ∼2 times higher at the dusk during southward IMF. During southward IMF, the proton intensity is significantly higher on the duskside compared to those during northward IMF (regions I and II, ion intensity is approximately 2 times higher).

9. There is a significant difference in the $O^+/H^+$ ratio between the two IMF directions at the dawn tail side (region VI). This means that acceleration, transport, and loss mechanisms are not the same for oxygen and proton ions for the two IMF directions. At the dayside the loss mechanism is similar for both oxygen ions and protons. However, during northward IMF the relative (ratio of intensity before (region III) and after (region IV)) loss of oxygen is more significant than of protons.

We summarize the findings on dawn-dusk asymmetries and the differences between quiet and disturbed magnetospheric conditions in Table 3. We also compare our results with those derived for the ion densities by *Maggiolo and Kistler* [2014].

## 4. Discussion

### 4.1. Dependence on Geomagnetic and Solar Wind Parameters

The direction of IMF regulates plasma transport in the magnetosphere [e.g., *Winglee*, 2000; *Welling and Ridley*, 2010; *Cully et al.*, 2003b]. The supply of ionospheric ions into the magnetotail plasma sheet depends on the IMF direction. During a southward directed IMF, a part of the outflowing ionospheric ions is captured on reconnecting field lines at the dayside and, due to enhanced convection, is transported to the central plasma sheet. There, the ions get trapped and are significantly accelerated, e.g., by tail reconnection or/and dipolarization. During a northward IMF, because of the weak convection, ionospheric ions mostly stay on the open field lines in the lobes and do not enter the plasma sheet. During periods of southward IMF, the abundance of ionospheric oxygen increases in the plasma sheet, where they can be further accelerated. One may therefore expect higher intensities of energetic oxygen fluxes during periods of southward IMF. A significant increase in oxygen ion intensities is indeed observed in the magnetotail regions I and VI (see section 3 and Figures 4 and 7) during southward IMF.

There is an increased possibility of reconnection events in the magnetotail at times when the magnetosphere is compressed, because the current sheet becomes thinner. However, the increase in SW $P_{dyn}$ shows a correlation with the increase in ion intensities only at the near-Earth duskside at least for our SW $P_{dyn}$ levels. Also, the ratio of oxygen and proton intensities is similar for both SW $P_{dyn}$ levels.

The most pronounced correlation is seen between ion intensities and geomagnetic activity. During substorms, the amount of oxygen increases significantly in the whole near-Earth magnetosphere. The clear dependence of the energetic ion distribution on the *AE* value can be due to the following effect. The partial disruption of the cross-tail electric current in the course of reconnection and/or current sheet disruption processes results in the formation of a substorm current wedge, reflected by high *AE* index values. These processes are followed by the generation of strong inductive electric fields which can be responsible for the effective ion energization. During periods of high *AE* we may expect an intensification of the ionospheric source triggered by precipitating particles during reconnection and/or current disruption. This provides an additional ion supply to the magnetotail. Quiet geomagnetic conditions, however, mean that there is no substorm current wedge formation and, therefore, no ion energization and no additional ionospheric source. Therefore, we observe quite low intensity levels for oxygen ions, see Figure 2.

The absence of a clear dependence of the energetic ion distributions on IMF and SW $P_{dyn}$ can be due to variety of acceleration processes and their delays in operation relative to those conditions.

### 4.2. Dawn-Dusk Asymmetries

The dawn-dusk asymmetry in the plasma sheet ion distribution can be caused by two general effects. One is gradient drift of the adiabatic ions toward the dusk flank. The other effect is when ions with gyroradii larger than the radius of curvature of the magnetic field lines in the current sheet experience nonadiabatic motion in the duskward direction and are accelerated by the dawn-dusk electric field (through the Speiser mechanism). The other effect occurs if there is a transient acceleration source. The ions are then subject to acceleration by inductive electric fields. It is worth to note that only inductive electric fields are capable of accelerating charged particles to energies exceeding the value of the cross-tail potential drop (> 100 keV) in the magnetotail. This mechanism is more complicated and depends on the spatial/temporal characteristics of the accelerating source. The simulations presented in previous studies [e.g., *Grigorenko et al.*, 2011, and references therein] showed that the dawn-dusk asymmetry in the energy distribution of 100 keV ions is not observed in the magnetotail if only inductive acceleration sources are operating. This is because under the influence of a strong time-dependent inductive electric field, a particle gains energy quickly and escapes the current sheet without a significant displacement in the dawn-dusk direction. The strength of the dawn-dusk asymmetry in the spatial distribution of energetic ions thus reflects the interplay between these two effects and inductive acceleration mechanisms. In this context we discuss in the following paragraph the features of the dawn-dusk asymmetry in the distributions of energetic protons and oxygen ions observed during the quiet and disturbed conditions.

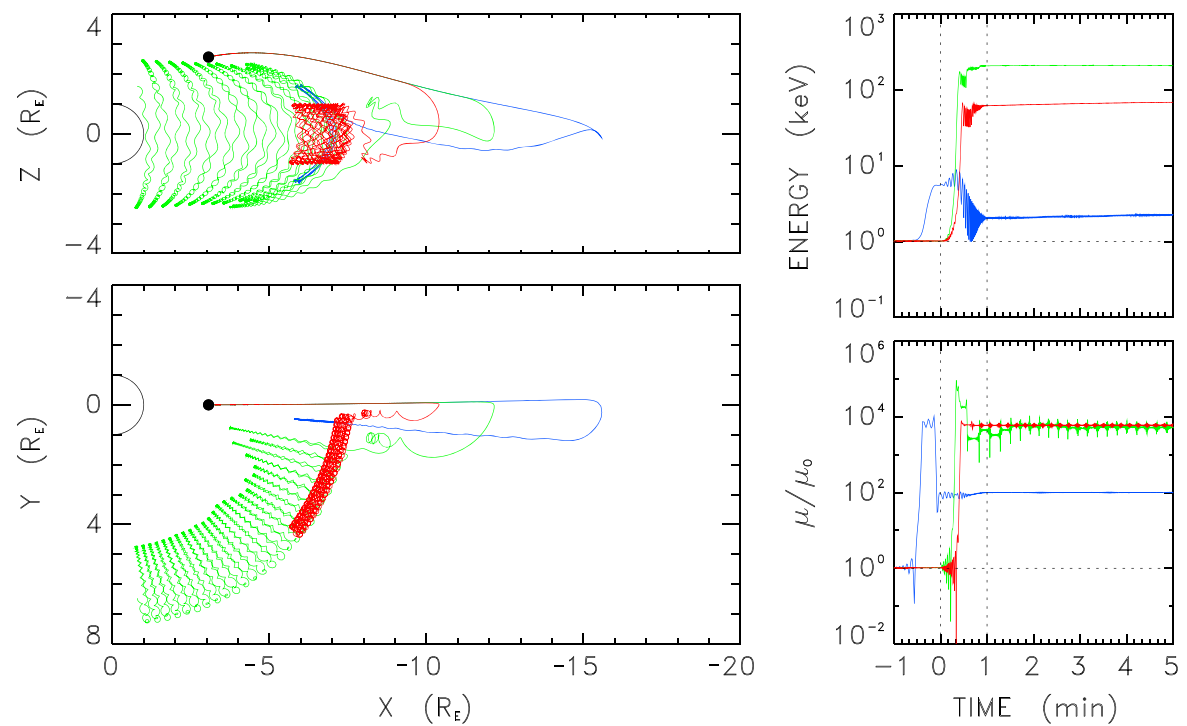

**Figure 8.** Trajectories of the 1 keV test protons are launched from the same position off equator (closed black circle in Northern Hemisphere) on midnight meridian from 4 $R_E$, 40° latitude with 160° pitch angle in (left) GSM *XZ* and *XY* planes. However, the protons are launched at distinct times before dipolarization and stopped at *t* = 5 min. (right) The dipolarization starts at *t* = 0 and terminates at *t* = 1 min as seen particles launched 3 min, 2 min, and 1.5 min before dipolarization, respectively. The figure shows how the induced electric field between 0 and 60 s can accelerate the particles (Figure 8, top right). Magnetic moments normalized to the initial value are also shown (Figure 8, bottom right).

Proton intensities do not show a significant dawn-dusk asymmetry during quiet geomagnetic periods. This confirm the importance of spatially localized inductive acceleration sources during quiet conditions. The value of the *AE* index does not necessary reflect the operation of transient localized dynamical processes in the magnetotail [e.g., *Grigorenko et al.*, 2013; *Luo et al.*, 2014].

During disturbed times, the duskward asymmetry is seen at the near-Earth nightside. This suggests the presence of two mechanisms: a strong duskward drift due to the increase of the magnetic field gradient in the near-Earth tail and strong inductive acceleration by nonstationary processes (magnetic dipolarization, turbulence, and transient reconnection) capable of energizing protons up to hundreds of keV [*Nosé et al.*, 2000; *Ono et al.*, 2009; *Luo et al.*, 2014]. A similar effect has been observed in the model by *Delcourt* [2002], where mass-selective ion energization occurs under the influence of a electric field induced by a time-varying magnetic field. The trajectories of protons subject to this acceleration are shown in Figure 8. This figure illustrates that the particles can get energized pretty fast without undergoing large distances in the dawn-dusk direction by the induced electric field between 0 and 1 min. This energization depends upon particle trajectory apex in the magnetotail as well as phasing with the dipolarizing field lines.

The oxygen ions at these energies are nonadiabatic even at the near-Earth magnetotail (3-D particle tracing modeling). They experience a duskward motion (along the dawn-dusk electric field) while convecting earthward. This leads to the dawn-dusk asymmetry during both quiet and disturbed times, in agreement with nightside observations. During quiet time periods, the presence of energetic protons (presumably produced through local inductive acceleration) and the relatively low intensity of the oxygen ions at the dawnside implies that the oxygen source is weak. During disturbed times, a dramatic increase of the oxygen intensity in the whole near-Earth plasma sheet (compared to protons) implies an additional source (ionosphere).

The duskward asymmetry of the energetic ion distribution is in agreement with the duskward distribution of the ion injections [*Gabrielse et al.*, 2014]. These injections were associated with localized dipolarizations which imply inductive acceleration.

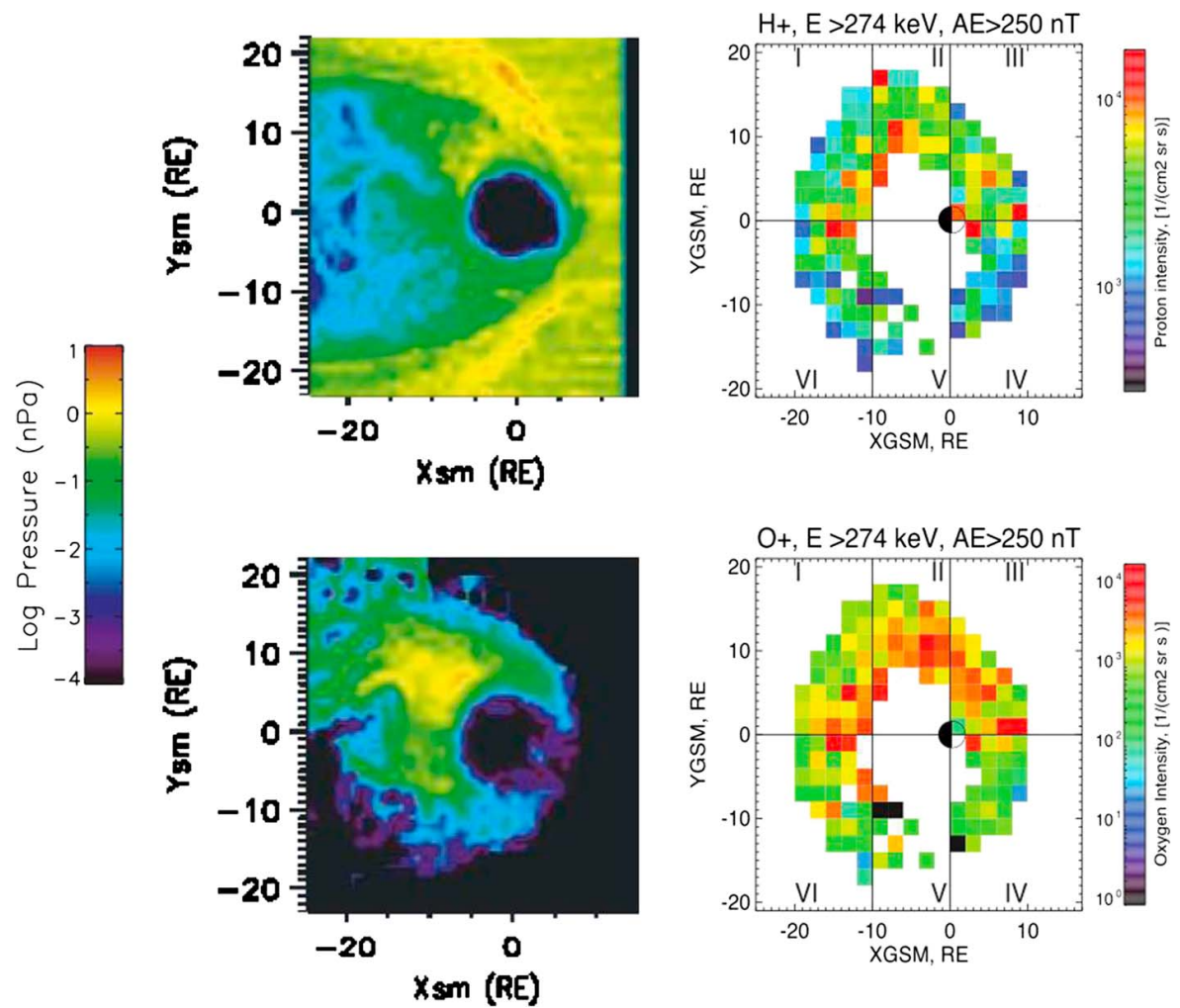

**Figure 9.** Comparison of simulations for the (top left) proton and (bottom left) oxygen pressures taken from *Fok et al.* [2006] with corresponding energetic intensities (top right) of the protons and (bottom right) of oxygen.

The >274 keV oxygen distribution pattern is similar to those seen in simulations by *Fok et al.* [2006] using the Lyon-Fedder-Mobarry MHD model [e.g., *Fedder et al.*, 1995] for substorm expansion phase, see comparison in Figure 9. We do not compare the growth phase as it is not clear to which *AE* index range it would correspond. For protons our map for geomagnetically disturbed times (Figure 9) is also in very good agreement with the simulations by *Fok et al.* [2006]. The dawn-dusk asymmetry is reflected by higher energetic ion intensities and by higher pressure at the near-Earth duskside. In the model by *Fok et al.* [2006] the energization of particles is due to inductive electric fields [*Delcourt*, 2002].

Also, the distribution of the relative energy density ($O^+/H^+$) for the $AE < 150$ nT simulated by *Winglee and Harnett* [2011] is similar to the corresponding map for the $O^+/H^+$ ratio, see comparison in Figure 10. They both show clear dawn-dusk asymmetry.

Considering asymmetries one has to compare the same physical quantities. Our study confirms previous observations of energetic particles and simulations of ion pressure and energy density which show a dawn-dusk asymmetry in the near-Earth magnetosphere. Observations of low-energy proton and oxygen densities [e.g., *Maggiolo and Kistler*, 2014] do not show such an asymmetry. Because of their lower energies (0–40 keV/*e*) the probability of the nonadiabatic effects in their dynamics is much smaller.

### 4.3. Losses

During disturbed times, the energetic ion intensity drops significantly between the postnoon and prenoon regions. This indicates particle sinks at these energies. Ions can be lost internally in the magnetosphere, for example, through charge exchange with neutral hydrogen at the geocorona or precipitation into the atmosphere through wave-particle interactions [*Kistler et al.*, 1989; *Jordanova et al.*, 1996]. However, these losses are not very likely to happen at our energies and locations, according to runs of the particle tracing model by *Delcourt et al.* [1990]. This can be due to leakage through the magnetopause in the dayside [*Keika et al.*, 2005]. This can happen when the magnetosphere is compressed. To provide some estimate for this loss, we assume that all ions have escaped through the duskside half of the dayside, a surface that we describe as an ellipsoid with axes $10R_E$, $14R_E$, and $9R_E$. If we take the values between regions III and IV for the quiet time and between regions II and IV for the disturbed time (as actually the intensities are the highest in region II for this time) from Table 2, the loss rate of the oxygen ions during quiet geomagnetic time is $\sim 9.1 \times 10^{22} \pm 2.3 \times 10^{22}$ s$^{-1}$ and during disturbed geomagnetic time $\sim 2.8 \times 10^{23} \pm 9.8 \times 10^{22}$ s$^{-1}$. For low and high levels of the SW $P_{dyn}$ we get $\sim 1.1 \times 10^{23} \pm 3.2 \times 10^{22}$ s$^{-1}$ and $\sim 2 \times 10^{23} \pm 5.7 \times 10^{22}$ s$^{-1}$, respectively. For northward and southward directions of IMF, the loss is $\sim 1.3 \times 10^{23} \pm 5.6 \times 10^{22}$ s$^{-1}$ and $\sim 1.2 \times 10^{23} \pm 7.2 \times 10^{22}$ s$^{-1}$. These numbers are comparable to the estimates of oxygen loss from the dayside plasma sheet by *Seki et al.* [2001] which is $\sim 10^{24}$ s$^{-1}$. These authors neglected the losses of the particles at energies higher than 17 keV. However, our estimations show that the energetic particles also play a role in the loss estimations (taking into account that we start from quite high energies, 274 keV). These values show that the absolute loss is controlled by the SW $P_{dyn}$ and depends on the geomagnetic activity.

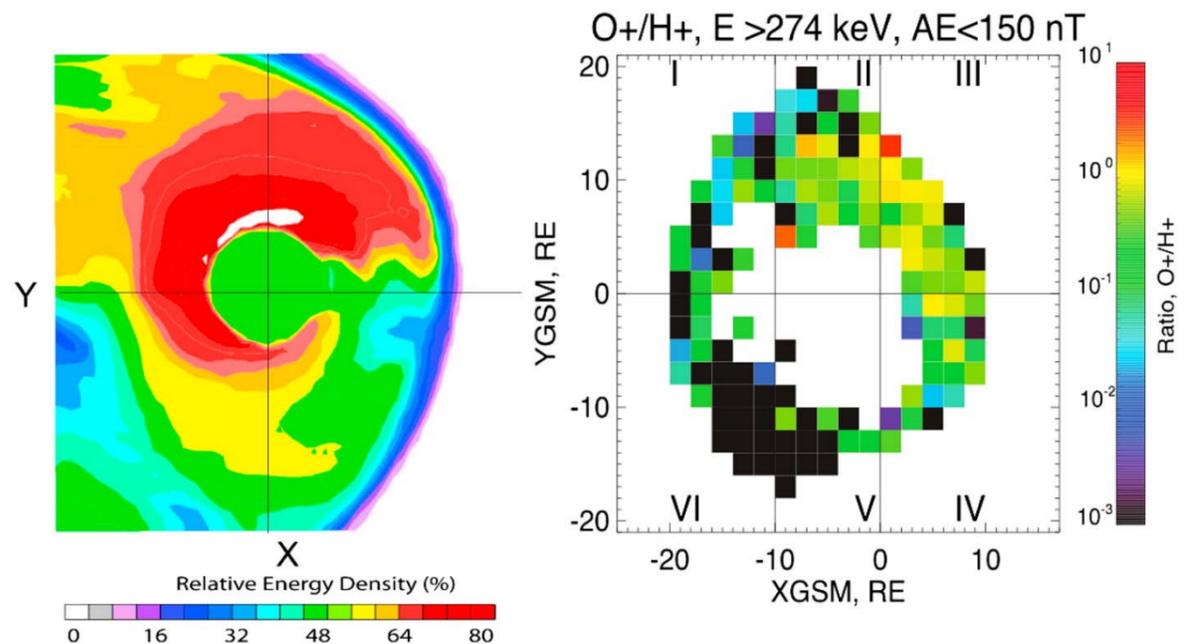

**Figure 10.** Comparison of simulations of relative energy density for $O^+/H^+$ at times when $AE < 150$ nT adapted from *Winglee and Harnett* [2011] with corresponding $O^+/H^+$ intensity ratio.

Another possible loss of energetic ions is energy diffusion which occurs during their drift around the Earth. Estimation of an interplay between the losses at the magnetopause and energy diffusion losses requires further studies.

It is also interesting that protons do not show any significant loss between the postnoon and prenoon regions for the northward IMF direction and also during low geomagnetic activity. Whether this is a gyroradius effect or, e.g., caused by entry through the magnetopause of the energetic upstream ions from the quasi-parallel bow shock needs to be investigated.

### 4.4. Pressure Estimations

The partial isotropic pressure can be derived using the following formula (see RAPID User Guide [*Daly and Kronberg*, 2014])

$$P(\text{nPa}) = 4\pi \frac{2}{3} 0.517 \times 10^{-8} \sqrt{m(\text{amu})} \sqrt{E(\text{keV})} J(\text{cm}^{-2}\text{sr}^{-1}\text{s}^{-1}), \quad (1)$$

where $m$ is the ion mass in atomic mass units (amu), $J$ is the integral intensity, and $E$ is the effective energy. The median intensity for the oxygen at energies >274 keV is 85 (cm$^{-2}$sr$^{-1}$s$^{-1}$) and for the protons 1300 (cm$^{-2}$sr$^{-1}$s$^{-1}$) during geomagnetically quiet times, 1300 and 2800 (cm$^{-2}$sr$^{-1}$s$^{-1}$) during geomagnetically disturbed times, respectively. This gives us $3.3 \times 10^{-4}$ nPa and $1.3 \times 10^{-3}$ nPa for the energetic oxygen and protons, respectively, during quiet times. The median proton pressure derived from CIS/CODIF observations is 0.17 nPa during quiet times. For the disturbed times the partial pressure of oxygen and protons at >274 keV is $5 \times 10^{-3}$ and $6 \times 10^{-3}$ nPa, respectively. This is insignificant compared to the median proton pressure 0.28 nPa derived from CIS. The two instruments may not be perfectly cross calibrated, but cross comparison between CIS and RAPID data reveals that intensities of RAPID ions may be higher than the CIS intensities but usually less than factor of 2 [*Kronberg et al.*, 2010]. The error in estimation of the partial pressure due to the different spectral shape of oxygen and protons is at most 20% and on average 7% [*Kronberg and Daly*, 2013]. Although the estimations are rough, they clearly show that the energetic ions at energies >274 keV do not significantly contribute (<2%) to the partial pressure produced by ions at 0–40 keV/$q$. We did the same estimations between 6 and 8$R_E$ as this region related to the ring current where energetic particles suppose to make significant contribution. However, the contribution of >274 keV ions is estimated to be <1% during disturbed time.

## 5. Summary

For the first time, based on 7 years of Cluster observations we established the distributions of energetic proton and oxygen intensities and their ratios at energies >274 keV in the near-Earth magnetosphere depending on the geomagnetic activity and SW activity. This information is important as an input for future verification of sources, acceleration mechanisms, transport, and ion losses.

From distributions we learn that

1. The distribution of ions in the magnetosphere is mass dependent. Energetic oxygen possesses the strongest spatial asymmetry depending on geomagnetic and solar wind activity.
2. The southward IMF leads to significantly stronger oxygen energization in the tail plasma sheet compared with the northward directed IMF.
3. The proton intensity shows significant increases at the duskside during disturbed geomagnetic conditions and at the near-Earth duskside during enhanced SW $P_{dyn}$ and southward IMF, implying there location of effective inductive acceleration mechanisms and a strong duskward drift due to the increase of the magnetic field gradient in the near-Earth tail.
4. The strongest changes of the ion intensities are associated with *AE* index and not the change of the IMF direction or SW $P_{dyn}$. This suggests that acceleration of ions is directly associated with inductive effects in the magnetotail during substorms.
5. We do observe the dawn-dusk asymmetry for the energetic ion intensities for disturbed conditions and for oxygen during quiet conditions. This is in agreement with previous observations of energetic particles and models which output the ion pressure and energy density and take into account ionospheric source. Most notably, such an asymmetry is not observed for the distributions of low-energy ion density [e.g., *Mouikis et al.*, 2010; *Maggiolo and Kistler*, 2014].
6. Higher losses of energetic ions are observed in the dayside plasma sheet under disturbed geomagnetic conditions and enhanced SW $P_{dyn}$.
7. Our observations are in agreement with models by, e.g., *Delcourt* [2002], *Fok et al.* [2006], *Welling and Ridley* [2010], and *Welling and Ridley* [2011].

**Acknowledgments**

We acknowledge the Deutsches Zentrum für Luft und Raumfahrt (DLR) for supporting the RAPID instrument at MPS under grant 50 OC 1401. E.E. Grigorenko thanks Russian Scientific Foundation (project 14-12-00824) for the financial support. We also thankful to the International Space Science Institute (ISSI) for their hospitality, and, for giving us the opportunity to gather the team on "Heavy ions: their dynamical impact on the magnetosphere" which allowed us productive collaboration. The authors are grateful for the use of the OMNI database for providing the *AE*, *Dst*, and solar wind parameters. The Cluster data can be found at CSA Archive: http://www.cosmos.esa.int/web/csa/.

Michael Balikhin thanks the reviewers for their assistance in evaluating this paper.